\documentclass[showpacs,preprintnumbers,amsmath,amssymb]{revtex4}
\usepackage{epsfig}

\newcommand{\calu}{{\cal U}}

\newcommand{\fith}{\widehat\varphi}
\newcommand{\gtil}{\widetilde{G}}
\newcommand{\gatil}{\widetilde{\Gamma}}

\newcommand{\barz}{\bar{z}}
\newcommand{\barg}{\overline{G}}
\newcommand{\barga}{\overline{\Gamma}}
\newcommand{\sech}{\hbox{sech}}
\def\ket{\rangle}        
\def\bra{\langle}        

\begin{document}
\setcounter{figure}{0}
\title{\bf Correlations around an interface}
\author{A. Bessa\footnote{abessa@if.ufrj.br}, C. A. A. de Carvalho\footnote{aragao@if.ufrj.br}, E. S. Fraga\footnote{fraga@if.ufrj.br}}
\address{Instituto de F\'\i sica, Universidade Federal do Rio de Janeiro\\
C.P.~68528, Rio de Janeiro, RJ 21941-972, Brasil}
\date{\today}


\begin{abstract}

We compute one-loop correlation functions for the fluctuations of an interface using a field theory model. We obtain them from Feynman diagrams drawn with a propagator which is the inverse of the Hamiltonian of a P\"oschl-Teller problem. We derive an expression for the propagator in terms of elementary functions, show that it corresponds to the usual spectral sum, and use it to calculate quantities such as the surface tension and interface profile in two and three spatial dimensions. The three-dimensional quantities are rederived in a simple, unified manner, whereas those in two dimensions extend the existing literature, and are applicable to thin films. In addition, we compute the one-loop self-energy, which may be extracted from experiment, or from Monte Carlo simulations. Our results may be applied in various scenarios, which include fluctuations around topological defects in cosmology, supersymmetric domain walls, Z(N) bubbles in QCD, domain walls in magnetic systems, interfaces separating Bose-Einstein condensates, and interfaces in binary liquid mixtures.
\end{abstract}

\pacs{11.10.Wx, 11.10.Kk, 11.27.+d, 64, 68}
\maketitle


\section{Introduction}
\label{introduction}

Many natural systems exhibit interfaces that separate regions of different physical characteristics. An interface in a binary liquid mixture separates its two components, a domain wall in a magnetic system separates its magnetic phases. Apart from those traditional examples, several systems of current interest can be viewed as different guises of that same physical situation. They include topological defects in cosmology\cite{Kibble}, supersymmetric domain walls\cite{peter}, Z(N) interfaces in thermal SU(N) gauge theories\cite{ZN-interfaces}, or different types of Bose-Einstein condensates\cite{bec}.

Recently, studies of quantum and statistical fluctuations around interfaces or domain walls have been the object of renewed attention. Such studies have concentrated on one-loop calculations which, technically, amount to computing fluctuation determinants around the interface (domain wall) background. They include computations around kink backgrounds in scalar theories in various dimensions\cite{yaffe}, as well as in supersymmetric models\cite{peter}. They use different methods, exploiting connections with special properties of determinants of differential operators\cite{yaffe}, with scattering data\cite{mit}, and with the spectrum of the operators\cite{peter}.

The calculations mentioned in the previous paragraph are restricted to vacuum bubbles, as will become clear in the sequel. In the present article, we will go beyond by computing {\em correlations} that involve one and two-point functions. Determinants, vacuum bubbles, and correlations will all be obtained from a semiclassical propagator that describes how the fluctuations of an interface (domain wall) evolve. Our method relies on a systematic semiclassical expansion around the given background, and has the semiclassical propagator as its essential ingredient. It not only serves as an alternative to the methods used in particle physics inspired applications \cite{peter,yaffe,mit}, but also provides a unified framework that extends those results to the computation of correlations, allows for connections with statistical mechanical systems (for which one-point functions have been obtained by other methods), and introduces calculations of two-point functions that lead to novel results.

We shall profit from the connections with statistical mechanical systems to present our method. Indeed, it is well-known\cite{jasnow=dombgreen, wallace} that interface fluctuations may be described by a scalar field theory model with a double-well potential. The model admits a classical solution, a kink profile depending on only one spatial (longitudinal) coordinate, which is associated with the mean-field configuration of the interface.

Fluctuations of the interface have been taken into account in $4-\epsilon$ dimensions, via renormalization group methods\cite{jasnow+rudnick=4-e,jasnow=dombgreen}, as well as directly in three-dimensions\cite{jasnow+rudnick=3d,munster=89}. As a result of their inclusion, calculations of the modified surface tension\cite{jasnow=dombgreen,brezin+feng,munster=89,munster=90,hoppe+munster}, and of the modified interface profile\cite{jasnow+rudnick=4-e,jasnow+rudnick=3d} were successfully carried out: the surface tension was computed up to two-loop order\cite{hoppe+munster}, leading to the prediction of universal ratios for an interface in the three-dimensional Ising model (which belongs to the same universality class of the scalar field model); the interface profile was computed up to one-loop order\cite{jasnow+rudnick=4-e,jasnow+rudnick=3d}, leading to a detailed comparison with experimental data\cite{wu+webb} for the reflectivity of binary liquid mixtures near a phase transition.

The three-dimensional calculations listed above can all be viewed as the result of computing the Feynman diagrams of a semiclassical expansion around the mean-field interface background\cite{aragao=TQFDW}. The diagrams involve a semiclassical propagator and semiclassical vertices\cite{annalsofphysics}: the former is the inverse of the Hamiltonian of a P\"oschl-Teller problem in one dimension; the latter include a background dependent cubic vertex, in addition to the quartic vertex of the double-well. 

In this paper, we use a closed analytic form for the semiclassical propagator that has been recently obtained\cite{aragao=TQFDW}, and show that it amounts to summing up the spectral representation for the inverse of the P\"oschl-Teller Hamiltonian. Indeed, from our expression, we recover the eigenvalues and eigenfunctions of that problem. We then use that compact expression to rederive previous results, to extend them to lower dimensions, and to compute two-point correlations up to one-loop.

The two-point vertices that we compute correspond to the self-energy at large relative distances (zero relative momenta). They depend on the position of those points relative to the interface, a consequence of the breaking of translational invariance. That self-energy is the sum of a mass (inverse correlation length) squared and a potential which reflects the influence of the interface on the fluctuation modes. We comment on how to compare our predictions for those quantities to experiments and simulations. 
 
The article is organized as follows: in Section II, we introduce the field theory model, and outline the derivation of the mean-field interface solution and of the semiclassical propagator, as well as the semiclassical expansions for the various generating functionals; in Section III, we compute the surface tension from the vacuum bubbles; in Section IV, we extract the interface profile from the graphs for the one-point Green function; in Section V, we obtain the self-energy from the graphs for the two-point vertex function; Section VI presents conclusions and suggestions for further work. Appendix A presents a detailed derivation of the semiclassical propagator, and explores its properties. Appendix B describes the renormalization procedure we have adopted.
   

\section{The field theory model}
\label{thefieldtheorymodel}

We consider the generating (partition) functional for a self-interacting scalar field theory model
\begin{equation}
\label{Z1}
Z[j]=\oint
[{\cal D}\varphi]\;\exp \bigg \{-\frac{S}{\lambda}+ \int \frac{j\,\varphi}{\sqrt{\lambda}}\,d^d x\bigg \}\; ,
\end{equation}
whose action functional in $d$ spatial dimensions is given by
\begin{equation}
S[\varphi]=\int d^d x\, \big[\frac{1}{2}(\nabla \varphi)^2+
\frac{1}{4!}(\varphi^2-\varphi_v^2)^2 \big]\;,
\end{equation}
where $\lambda$ is a dimensional coupling, $\varphi_{v}$ is the vacuum value of the field $\varphi$, and $j(x)$ is an external current.

The model may be used to describe an interface and its fluctuations: for instance, we may associate the scalar field to the difference in concentration of the phases (components) of a binary system, as in a binary liquid mixture. Similarly, we may relate it to an Ising-like spin. The phases of the liquid mixture are described by the two degenerate vacua $\pm\varphi_{v}$ of the model. Analogously, the two homogeneous configurations where the spins are either all up, or all down, may likewise be described by those two vacua. The mean-field interface will emerge as a classical solution of the equation of motion of the model. Its fluctuations will be captured by a semiclassical expansion around that solution.

The classical solution of interest is the well-known kink profile, which depends on only one longitudinal coordinate $z$,
\begin{subequations}
\begin{equation}
\fith(z) = \pm \varphi_{v} \tanh \theta(z) \;\; ,
\end{equation} 
\begin{equation}
\theta(z)= \frac{M}{2} \,(z-\barz)\; ,
\end{equation} 
\end{subequations}
where $M \equiv \varphi_{v}/\sqrt{3}$. The solution breaks translational invariance along the longitudinal direction as it depends on a position $\barz$, the point where it vanishes, which can be identified with the position of the kink. For two and three dimensions, the dependence of $\barz$ on the transverse coordinate(s) characterizes the interface.

The semiclassical expansion includes fluctuations of the interface in a systematic way\cite{aragao=TQFDW,annalsofphysics}. Setting
\begin{equation}
\label{expansion}
\varphi(x)=\fith(z) + \lambda^{1/2}\eta(x)\;,
\end{equation}
we expand the action functional around the classical solution, regarded as a mean-field profile to be modified by the fluctuations, and perform a saddle-point integration to obtain the generating functional of Eq.\eqref{Z1} in the form of an infinite (semiclassical) series.

The quadratic term in the functional expansion of the action defines the semiclassical propagator $G(x,x')$ around the kink background 
\begin{equation}\label{DiffG(x,x')=delta}
\big [-\nabla^{2} + \frac{1}{2}(\fith^2 -M^2) \big ]\, G(x,x') = \delta^{d}(x-x')\; .
\end{equation}
Using the expression for $\fith$, and Fourier transforming the transverse coordinates, leads to
\begin{equation}\label{DiffG(k;z,z')=delta}
\big [-\partial^{2}_{z} + {\vec{k}}^{2}+ M^2 -\frac{3}{2} M^2 \sech^{2}\theta\big ] \barg(\vec{k};z,z') = \delta(z-z')\; ,
\end{equation}
where $\vec{k}$ is the transverse momentum, and $\barg$ is a hybrid momentum-position propagator.

$\barg$ can be viewed as the inverse of the Hamiltonian for a Schr\"odinger problem in one dimension with a P\"oschl-Teller potential $U(\theta)\equiv  \vec{k}^{2}+ M^2 -(3/2)M^2\sech^{2}\theta$. We may obtain an expression for $\barg$ from two linearly independent solutions of the homogeneous version of Eq.\eqref{DiffG(k;z,z')=delta}. It is a hypergeometric equation, but it so happens\cite{aragao=TQFDW} that its two independent solutions are hypergeometric series that terminate, as shown in Appendix A. Therefore, we end up with an expression for $\barg$ which can be written in terms of elementary functions. Using the dimensionless quantities $\vec{\kappa} \equiv 2\vec{k}/M$, $u\equiv (1-\tanh \theta)/2$, and $b\equiv \sqrt{4+\vec{\kappa}^2}$
\begin{subequations}
\begin{equation}\label{propsc(b,u)}
\barg = \frac{2}{M} \bigg \{ \frac{1}{2b} \big[\phi(b,u)\phi(-b,u')\Theta(u'-u)+ \phi(-b,u)\phi(b,u')\Theta(u-u')\big ] \bigg \}\; , 
\end{equation}
\begin{equation}\label{fi(b,u)=...}
\phi(b,u) \equiv \bigg(\frac{u}{1-u} \bigg)^{b/2}\; \bigg [1 - \frac{6u}{b+1}+\frac{12u^2}{(b+1)(b+2)} \bigg ] \equiv \bigg(\frac{u}{1-u} \bigg)^{b/2}\, f(b,u)\;.
\end{equation}
\end{subequations}
In Appendix A, we show that the above form is equivalent to the usual spectral sum over eigenmodes. In fact, we use it to derive the eigenvalues and eigenfunctions of the Schr\"odinger problem. 

The cubic and quartic terms in the functional expansion of the action define the vertices of the semiclassical series
\begin{equation}\label{v3andv4}
V_{3} = \frac{\lambda^{1/2}}{3!} \fith \eta^{3}\; ;\;\; \;\; V_{4} = \frac{\lambda}{4!}\eta^{4}\;\; .
\end{equation}   
The cubic vertex involves the kink background. Saddle-point integration implies expanding the exponential of the cubic and quartic parts of the action in a power series, and then performing the resulting functional integrals, which reduce to powers of the fluctuation times a Gaussian, leading to a series for the generating functional.

The translational invariance of the model requires the introduction of a collective coordinate. The background solution explicitly breaks the invariance along the longitudinal direction (it is centered on $\barz$). However, as there are solutions for any value of $\barz$, translational invariance must be restored in the calculation by adding over all possible values of $\barz$. Indeed, translational invariance manifests itself through the appearance of a zero-energy eigenmode of the fluctuation kernel of Eq.\eqref{DiffG(k;z,z')=delta}. Physically, it costs zero energy to infinitesimally translate a classical solution. The restoration of the symmetry is accomplished by means of the Faddeev-Popov procedure, and yields a Jacobian. We trade integration over the zero-mode subspace for integration over the collective coordinate $\barz$. 

The generating functional, after the introduction of the collective coordinate, becomes
\begin{equation}\label{Z2}
Z[j]= \int_{-L/2}^{L/2} \frac{d\barz}{L} \bigg (\frac{\widehat{S}}{2\pi \lambda} \bigg )\;  (\Delta')^{-1/2}\, {\cal P}\bigg [\fith,\frac{\partial}{\partial j} \,z '[j]\bigg ] \;\exp \bigg \{-\frac{\widehat{S}}{\lambda}+ \int \frac{j\,\fith}{\sqrt{\lambda}}\,d^d x\bigg \}\; .
\end{equation}
We have introduced a longitudinal infrared cutoff $L$, the longitudinal size of the system; the $(\widehat{S}/\lambda)^{1/2}$ factor comes from the Jacobian, while the $(2\pi)^{-1/2}$ comes from the functional measure;  $\Delta'$ is the determinant of $[G']^{-1}$, with $G'$ denoting the semiclassical propagator with the zero-mode subspace excluded; $z'[j] \equiv \exp \{\frac{1}{2}\int{j(x)G'(x,y)j(y)d^d x d^d y} \}$; and ${\cal P}[\fith,\eta]$ is defined by the power series expansion of the exponential in
\begin{subequations}\label{pertubation[phi,eta]}
\begin{equation}
{\cal P}[\fith,\eta]\equiv \bigg (1-\frac{\lambda^{1/2}}{\widehat{S}} \int{\calu '[\fith]\eta}\bigg )\, e^{-\delta S}\;\; ,
\end{equation}
\begin{equation}
\delta S \equiv \int{\big (V_{3}+V_{4}}\big )\,d^d x\; . 
\end{equation}
\end{subequations}
$\calu '[\fith]$ is the derivative of the double-well potential $\calu \equiv (\varphi^2-\varphi_{v}^{2})^{2}/4!$ with respect to $\varphi$, computed at the kink solution; $V_{3}$ and $V_{4}$ are the vertices of Eq.\eqref{v3andv4}. The Faddeev-Popov Jacobian expanded around $\fith$ yields the combination $(\widehat{S}/\lambda)^{1/2}{\cal P}$ appearing in Eq.\eqref{Z2}.

Connected correlation functions can be derived from the free energy functional $F[j] = -\lim_{L\rightarrow \infty}\{\log Z[j]\}$ by functional differentiation with repect to $j(x)$. A Legendre transform leads to the effective action (Gibbs) functional $\Gamma[\phi] = F[j]+\int{j\,\phi/\sqrt{\lambda}\, d^d x}$, with $\phi(x)\equiv \bra \varphi(x)\ket$ denoting the expectation value of the field $\varphi$. Its functional derivatives with respect to $\phi(x)$ lead to one-particle irreducible $(1PI)$ vertex functions. Connected correlations and $1PI$ vertex functions can all be expressed in the usual language of Feynman diagrams, as in the explicit examples of the forthcoming sections.  We note that a semiclassical expansion around $\fith\;\; (\varphi_{v})$ leads naturally to correlations and vertex functions in the interface (vacuum) sector, as we take functional derivatives at $j=0$, for the former, and at $\phi = \fith \;\;(\varphi_{v})$, for the latter.

Finally, renormalization is required to connect the various correlation functions to physical parameters. As we are interested in comparing correlations in the presence of an interface with those of the homogeneous (vacuum) phase ({\it i.e.}, with bulk values), we will use counterterms computed in the vacuum sector. The diagrams in the kink and in the vacuum sector will then be related to the bulk physical parameters by means of a standard renormalization procedure, which is outlined in Appendix B.

\section{The surface tension}
\label{thesurfacetension}

The surface tension for the interface may be obtained from the difference between the free energy functionals at zero external current in the interface and vacuum sectors
\begin{equation}
\sigma \equiv \lim_{A \rightarrow \infty}\frac{\Delta F[0]}{A} = \lim_{A\rightarrow \infty}\frac{F_{i}[0]-F_{v}[0]}{A}\;\; ,
\end{equation}
where $A$ denotes the \lq\lq area\rq\rq \, spanned by the transverse direction(s). Its leading (zero-loop) value is given by the difference of the classical actions in the two sectors $\sigma_{0} = \lim_{A \rightarrow \infty}(\widehat{S}/\lambda A) =2M^3/\lambda$, which is independent of the space dimension.

Up to one-loop order, the free energy and effective action functionals coincide\cite{amit}, so we have
\begin{equation}
\Delta F[0]= \Gamma[\fith] -\Gamma[\varphi_{v}] = \frac{\widehat{S}}{\lambda} + \frac{1}{2} \log \bigg (\frac{\Delta '  }{\Delta_{v}}\frac{2\pi \lambda}{\widehat{S}} \bigg) \;\; ,
\end{equation}
where $\Delta_{v}$ is the (free) determinant of the quadratic fluctuation kernel $G_{v}^{-1}$ in the vacuum sector. Neglecting terms that vanish as $(\log A/A)$, we have
\begin{equation}
\sigma = \sigma_{0}+\sigma_{1} = \sigma_{0}+ \lim_{A \rightarrow \infty}\bigg (\frac{1}{2A}\log\frac{\Delta'}{\Delta_{v}} \bigg )\;\; . 
\end{equation}

The $\sigma_{1}$ contribution can be computed from the semiclassical propagator of Appendix A, as shown in reference \cite{aragao=TQFDW}. Indeed,
\begin{equation}\label{sigma1=int(ln[Delta/Deltav])}
\sigma_{1} = \frac{1}{2} \int{\frac{d^{d-1}k}{(2\pi)^{d-1}} \log \frac{\Delta(k)}{\Delta_{v}(k)}}\;\; ,
\end{equation}
where $\Delta(k)$ and $\Delta_{v}(k)$ are the determinants of $[\barg(k)]^{-1}$ and $[\barg_{v}(k)]^{-1}$, and we sum over transverse momenta. Exclusion of the zero-mode for $k=0$ is guaranteed by writing
\begin{equation}
\log \frac{\Delta(k)}{\Delta_{v}(k)} = \log \kappa^2(k) + \log \frac{\Delta '(k)}{\Delta '(0)} - \log \frac{\Delta_{v}(k)}{\Delta_{v}(0)}+\log \frac{\Delta '(0)}{\Delta_{v}(0)}\;\; .
\end{equation}
The second term on the rhs is expressible in terms of the semiclassical propagator 
\begin{equation}
\log \frac{\Delta '(k)}{\Delta '(0)} = \int_{0}^{\kappa^2} ds \,\int_{-\infty}^{\infty}{dz \, \barg'(\sqrt{s};z,z)}\;\; ,
\end{equation}
whereas the third is given by a similar expression with $\barg'$ replaced by $G_{v}$. Then
\begin{equation}\label{log(delta'/delta')-log(deltav'/deltav')}
\log \frac{\Delta '(k)}{\Delta '(0)}\, -\, \log \frac{\Delta_{v} '(k)}{\Delta_{v} '(0)} = \log\bigg [\frac{b(k)-1}{b(k)+1} \bigg ]\,- 2\log [b(k)+2]\,+\log 48\;\; .
\end{equation}
The fourth term may be computed following reference \cite{rajaraman} and yields $\log [\Delta^{\prime}(0)/\Delta_{v}(0)]=-\log 48$. Using \eqref{sigma1=int(ln[Delta/Deltav])} and \eqref{log(delta'/delta')-log(deltav'/deltav')}, we derive the unrenormalized expression 
\begin{equation}\label{sigma1=int(log(bk-1)(bk-2)/(bk+1)(bk+2)} 
\sigma_{1} = \frac{1}{2}\, \int\frac{d^{d-1} k}{(2\pi)^{d-1}}\, \log \bigg \{ \frac{[b(k)-1]\,[b(k)-2]}{[b(k)+1]\,[b(k)+2]} \bigg \}\;\; ,  
\end{equation}  
which is valid for $k>0$.  
 
In $d=2$ and $d=3$, expression \eqref{sigma1=int(log(bk-1)(bk-2)/(bk+1)(bk+2)} needs to be regularized and renormalized. We use a cutoff in transverse momentum space to regularize: in $d=2$, we integrate over the interval $[-\Lambda,\Lambda]$, whereas, in $d=3$, we integrate over a disk of radius $\Lambda$. Neglecting terms that vanish as $1/\Lambda$, we obtain  
\begin{subequations} 
\begin{align}
&\sigma_{1}= -\frac{3M}{2\pi}\, \log \bigg (\frac{2\Lambda}{M} \bigg ) + \bigg (\frac{1}{4\sqrt{3}} -\frac{3}{2\pi}\bigg )\;\;, \;\; d=2\,,\\ 
&\sigma_{1}= -\frac{3M\Lambda}{4\pi}+ \bigg (\frac{3}{8\pi} -\frac{3\log 3}{32\pi}\bigg )\,M^2\;\;, \;\; d=3\; . 
\end{align} 
\end{subequations}

The renormalization procedure is described in Appendix B. We adopt renormalization conditions at zero momenta 
\begin{equation}\label{gammaR(1)=0,gammaaR(2)=M2} 
\barga_{R}^{(1)}(0)\,=\,0\;\;\; ; \hspace{1cm}\barga_{R}^{(2)}(0)\,=\,M^2\;\; , 
\end{equation} 
which specify the Fourier transformed one and two-point $1PI$ vertex functions. They lead to the renormalized result 
\begin{subequations}
\begin{align} 
&\sigma_{1R}=  \bigg (\frac{1}{4\sqrt{3}} -\frac{3}{2\pi}\bigg )\;\;, \;\; d=2\,,\\ 
&\sigma_{1R}= \frac{3}{32\pi}\,\big (\log 3 -4\big )\,M^2\;\;, \;\; d=3\; . 
\end{align} 
\end{subequations}
For $d=2$, the result coincides with the correction to the kink mass obtained previously\cite{rajaraman}. The results for $d=3$ are shown, in Appendix B, to coincide with those in the literature\cite{jasnow=dombgreen,brezin+feng,munster=89,munster=90,peter}. 
 
\begin{figure}[h]
\vspace{5mm}
\centerline{\epsfig{figure=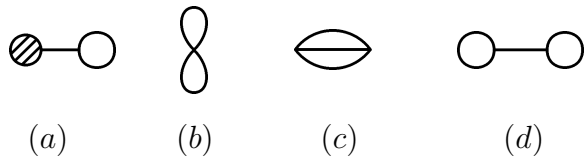,width=7cm}}
\caption{Diagrams which contribute to the surface tension at two-loop order.}\label{fig2loop}
\vspace{5mm}
\end{figure}

We could go beyond one-loop in our expansion by including contributions which are first-order in the cubic and quartic semiclassical vertices. Using equations \eqref{Z2} and \eqref{pertubation[phi,eta]} to compute $F[0]$ to that order, we obtain the two-loop Feynman diagrams and the Jacobian contribution depicted in Figure \ref{fig2loop}. Such diagrams were computed in reference \cite{hoppe+munster}, using the spectral sum representation for the semiclassical propagator. We believe that our (resummed) expression for the propagator will confirm their results, and simplify the calculation, but we shall postpone that verification for a future publication, and concentrate on one-loop correlations in the present article. As the two-loop results have been compared to those obtained from Monte Carlo simulations \cite{montecarlo}, it is important to have an independent check. Presumably, the calculation will be more direct if one makes use of the compact expression for the resummed propagator. 
 
\section{The interface profile} 

\begin{figure}[h]
\vspace{5mm}
\centerline{\epsfig{figure=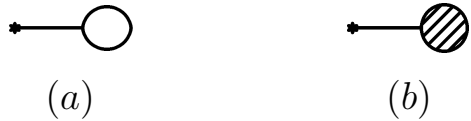,width=7cm}}
\caption{Contributions to the interface profile. Diagram $(b)$ represents the Jacobian term. }\label{figperfil}
\vspace{5mm}
\end{figure}

The interface profile is given by the expectation value of the order parameter $\phi(x)=\bra \varphi (x) \ket$, which we compute from the first derivative of $F[j]$ with respect to the external current $j(x)$, at $j=0$. The calculation was carried out in the kink sector using the semiclassical expansion around $\fith$. We may express the correction to the interface profile in terms of the Feynman diagrams of Figure \ref{figperfil}. As we shall show, this is equivalent to solving a one-loop corrected equation for $\phi$, as was done in references \cite{jasnow=dombgreen,jasnow+rudnick=4-e,jasnow+rudnick=3d}. It can be calculated in a much simpler and compact way using our semiclassical propagator. 

Before proceeding, we return to the discussion of translational invariance. Inspection of the diagrams in Figure \ref{figperfil} shows that diagram (a) is ultraviolet divergent in $d=3$ if we compute it with the semiclassical propagator in the subspace orthogonal to the zero-mode. Indeed, the excluded zero-mode contribution behaves as  
\begin{equation}
\int_{\Lambda}\frac{d^2 k}{(2\pi)^2}\int_{-\infty}^{\infty}dz\,\frac{\overline{\eta}_{0}^{2}(z)}{\vec{k}^{2}} \,\propto\, \log \frac{2\Lambda}{M}\;\; ,
\end{equation}
which accounts for the divergence. The problem does not exist in lower dimensions, suggesting that $d=3$ is a marginal dimension\cite{jasnow=dombgreen}. Were we to work with the full propagator, no ultraviolet problem would occur, but we would have an infrared problem for vanishing $\vec{k}$.

Following the interpretation of references \cite{jasnow=dombgreen,jasnow+rudnick=4-e,jasnow+rudnick=3d} for the case of binary liquid mixtures, we take that to indicate the instability of a translationally invariant interface in three dimensions. To cope with this problem, we explicitly break translational invariance by introducing a small mass $\mu$ in the zero-mode subspace. Physically, in three-dimensional binary liquid mixtures that can be attributed to the action of a gravitational field, or some other pinning effect, and is negligible in the other subspaces ($\mu \ll M$). In other physical applications, the fact that the interfaces or domain walls have their positions pinned down by some external effect which breaks translational invariance will be encoded in the dependence of $\mu$ on whatever parameter characterizes that pinning effect. 

Breaking translational invariance has the following implications for our calculation: i) we no longer need to work in the subspace orthogonal to the zero-mode, so that no Jacobian will emerge; ii) for the semiclassical propagator, we use the expression in the subspace orthogonal to the zero-mode, added to a zero-mode part with mass $\mu$
\begin{equation}
\barg(\vec{k};z,z') = \barg^{\prime}(\vec{k};z,z') + \frac{M}{2}\,\frac{\overline{\eta}_{0}(z)\,\overline{\eta}_{0}(z')}{\mu^2}\;,
\end{equation}
where $\overline{\eta}_{0}$ is the normalized zero eigenmode shown in Appendix A.

\begin{figure}[h]
\vspace{5mm}
\centerline{\epsfig{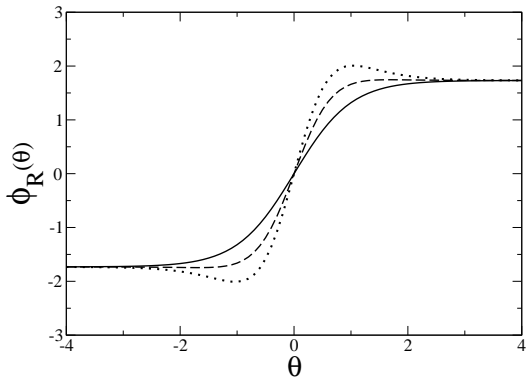}}
\caption{Profile diagrams in $d=2$ for $\lambda=0.1$. The solid line is the kink configuration. The dashed curve correspond to $\mu/M =0.01$ and the dotted one to $\mu/M =0.001$.}\label{profiled2} 
\vspace{5mm}
\end{figure}

The preceding paragraph implies that the only diagram to be considered in one-loop order is the first diagram of Figure \ref{figperfil}. We compute it by integrating over longitudinal coordinates and transverse momenta. With the renormalization conditions defined in \eqref{gammaR(1)=0,gammaaR(2)=M2}, we derive the renormalized profile
\begin{align}\label{profileform}
\phi_{R}(\theta)= \,\sqrt{3} M\,\tanh \theta -\lambda \, \big ( \,\alpha\, \theta \,\sech^2 \theta \;+\; \beta \,\tanh \theta \, \sech^2 \theta\,\big )\;\; , \;\; d=2 \hbox{ or } 3\;. 
\end{align}
The profile has the functional form above, and only the coefficients change with the spatial dimension. That is a direct consequence of the form of the semiclassical propagator. The computed coefficients are $\alpha = (2\pi-3\sqrt{3})/(16\pi M)$, $\beta =-(6\sqrt{3}+4\pi-3\sqrt{3}\pi M/\mu)/(24\pi M)$, for $d=2$, and $\alpha = 3\sqrt{3}(\log 3 -1)/(32\pi)$, $\beta = (\sqrt{3}/16\pi)\times \log(4M/3\mu)$, for $d=3$. It should be remarked that, for $d=3$, our calculation will coincide with that of reference \cite{jasnow+rudnick=3d} for a judicious choice of renormalization conditions. The latter reference confronted its findings with experimental results\cite{wu+webb}, being compatible with the data available at the time. Our new results for $d=2$, depicted in Figure \ref{profiled2}, illustrate the dependence of the profile on the ratio $\mu/M$. Obviously, the lower that ratio, the more striking the effect will be. Those results can be tested experimentally by studying the interface of thin films of binary mixtures\cite{binder}, for instance. More recent applications, such as the ones involving Bose-Einstein condensates, could also be used in experimental checks in $d=2$ and $d=3$\cite{bnu}.

As we have already remarked, it is only in $d=3$ that we are forced to break translational invariance. In $d=2$, a translationally invariant mean-field interface solution is stable, so that we may compute its fluctuations by using $\barg^{\prime}$, and including diagram (b) of Figure \ref{figperfil}. However, in the limit of large transverse ``area´´ (in the present case, a length), that contribution is negligible. The result we obtain has the functional form presented in \eqref{profileform} with $\alpha = (2\pi-3\sqrt{3})/(16\pi M)$ and $\beta =-(6\sqrt{3}+4\pi)/(24\pi M)$ (which is equivalent to taking $\mu\to\infty$ in the expression for $\beta$). That is to be compared with the profile obtained previously, without translational invariance.

\section{The two-point correlations}

We may take a second derivative of the free energy functional $F$ with respect to the external current $j$ to obtain the connected two-point function $G_{c}^{(2)}$. That leads to the Feynman diagrams of Figure \ref{figG(y1,y2)}. Just as in the previous section, one has to omit the diagram coming from the Jacobian, whenever translational invariance is broken. 

\begin{figure}[h]
\vspace{0.5cm}
\centerline{\epsfig{figure=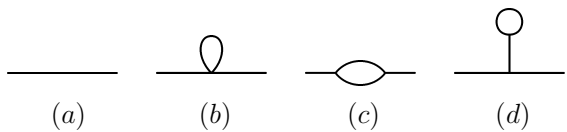,width=7cm}}
\caption{Diagrams which contribute to the two-point function up to one-loop order.}\label{figG(y1,y2)}
\end{figure}

The calculation of the two-point function $G_{c}^{(2)}$ up to one-loop order is rather involved, so we shall postpone it for a future publication. Instead, we will concentrate on mean-field results obtained from its lowest order expression, which is given by our semiclassical propagator, and on one-loop corrections to $\Gamma^{(2)}$. 

As our classical interface profile depends on a collective coordinate $\barz$, the hybrid Fourier transform $\barg(\vec{k};z,z')$ depends on both $z$ and $z'$, not just on their difference. Introducing center-of-mass and relative coordinates
\begin{subequations}
\begin{align}
&R=\frac{z+z'}{2}\;\; ,\\
&\rho = z - z'\;\; , 
\end{align}
\end{subequations}
we may reexpress $\barg$ in terms of $R$ and $\rho$. The expression has a particularly simple form for $\vec{k}=\vec{0}$  
\begin{equation}\label{G(R,rho,0)}
\barg(R,\rho,\vec{0}) = \frac{e^{M\rho}(6(M/\mu)^2-4-3M\rho)+ 8e^{\frac{M}{2}\rho}\cosh(MR)+\cosh(2MR)}{M \big (1+e^{M\rho} +2e^{M\rho/2}\cosh (MR)\big)^{2}}\;\Theta(\rho)\hspace{2mm}+\hspace{2mm}(\rho \;\leftrightarrow \;-\rho)
\end{equation}

If we now perform a Fourier transform in the $\rho$-coordinate, we obtain a function $\gtil$ depending on $R$, $k_{\rho}$ and $\vec{k}$. Setting $k_{\rho}=0$ and $\vec{k}=\vec{0}$ amounts to integrating over all relative coordinates. 

One may define a susceptibility $\chi$ as 
\begin{equation}\label{Suscep}
\chi=\lim_{L\rightarrow \infty}\frac{1}{L}\int_{-L/2}^{L/2}\,dR\, \gtil(R;0,\vec{0})\;\; .
\end{equation}
Likewise, one may exclude the lowest mode and define $\chi'$ using $\gtil'$ in the previous formula. For both cases, we obtain $\chi=1/M^2$, just as in the vacuum sector. That is a consequence of the fact that the two lowest modes are localized, whereas the continuum ones behave asymptotically as plane waves of mass $M$. Accordingly, the exponential decay of our propagator as the relative distance becomes large is of the form $exp(-M\sqrt{\rho^2+\rho_T^2})$, where $\vec{\rho}_T$ is the relative transverse coordinate. The correlation length is thus set by $1/M$, being independent of $R$, even when we give a small mass $\mu$ to the lowest mode. In principle, the susceptibility and the correlation length could depend on the position of the two points with respect to the interface, i.e., on $R$. However, the phases on either side of the interface are degenerate in our model. They have the same correlation length $1/M$. It is then natural that $M$ should set the scale. A model wherein the co-existing phases could have different masses (inverse correlation lengths) would probably lead to position-dependent quantities.

\begin{figure}[h]
\vspace{1.5cm}
\centerline{\epsfig{figure=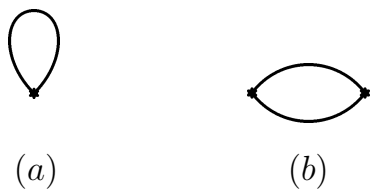,width=7cm}}
\caption{Diagrams which contribute to the self-energy up to one-loop order.}\label{figautoenergia}
\end{figure}

The calculation of the two-point vertex $\Gamma^{(2)}$ up to one-loop order involves fewer integrals than that of its inverse $G_{c}^{(2)}$. Setting 
\begin{equation}\label{Gammak2(x,x')}
\Gamma^{(2)}(x,x')\,\equiv\,{[G_{c}^{(2)}]}^{-1}(x,x') = G^{-1}(x,x')\,+\, \Sigma(x,x')\;\; ,
\end{equation}
where the first term on the rhs is the inverse of the semiclassical propagator, then $\Sigma(x,x')$ will be the contribution to the self-energy from the kink sector, and can be identified with two of the diagrams of Figure \ref{figG(y1,y2)} without external legs, as shown in Figure \ref{figautoenergia} (diagram $(d)$ is obtained as a combination of $\Gamma^{(1)}$ and $\Gamma^{(3)}$ with $G^{-1}$). 

We should stress that the semiclassical propagator itself already gives a contribution $\Sigma_{sc}$ to the self-energy when compared with the vacuum sector
\begin{equation}\label{Gamma-1(x,x')}
G^{-1}(x,x')= G_{v}^{-1}(x,x')+ \Sigma_{sc}(x,x')\;\;\; ,
\end{equation}
where $G_{v}^{-1}$ is the inverse of the free propagator. Equation \eqref{Gamma-1(x,x')} defines $\Sigma_{sc}$\,, just as equation \eqref{Gammak2(x,x')} defines $\Sigma$. Furthermore, the diagrams of Figure \ref{figautoenergia}, when computed with the free propagator, and with $\varphi_{v}$ instead of $\fith$ at the cubic vertices, will yield $\Sigma_{v}(x,x')$, defined by
\begin{equation}\label{Gammav2(x,x')}
\Gamma_{v}^{(2)}(x,x') \equiv {[G_{v}^{(2)}]}^{-1}(x,x') = G_{v}^{-1}(x,x') + \Sigma_{v}(x,x')\;\; .
\end{equation} 
Equations \eqref{Gammak2(x,x')}, \eqref{Gamma-1(x,x')} and \eqref{Gammav2(x,x')} lead to
\begin{equation}
\Delta\Gamma^{(2)}(x,x')= \Gamma^{(2)}(x,x')-\Gamma_{v}^{(2)}(x,x') = \Sigma_{sc}(x,x')+ \Sigma(x,x')-\Sigma_{v}(x,x') = \Delta \Sigma\; ,
\end{equation} 
the difference in self-energy between kink and vacuum sectors.

As in the case of the two-point function, the hybrid Fourier transform $\Delta \overline{\Sigma}(\vec{k};z,z')$ will depend on both $z$ and $z'$, not just on their difference, so that we may reexpress $\Delta \overline{\Sigma}$ in terms of $R$ and $\rho$. Performing, as before, a Fourier transform in the $\rho$-coordinate, we obtain a function $\Delta \widetilde{\Sigma}$ depending on $R$, $k_{\rho}$ and $\vec{k}$
\begin{equation}
\Delta \widetilde{\Sigma}(R;k_{\rho},\vec{k}) = \gatil^{(2)}(R;k_{\rho},\vec{k})-\gatil^{(2)}_{v}(R;k_{\rho},\vec{k})\;\; .
\end{equation}
In order to understand the physical meaning of the self-energy in the present situation, it is instructive to consider the lowest order term defined in Eq.(\ref{Gamma-1(x,x')}). From Eq.(\ref{DiffG(k;z,z')=delta}), we have
\begin{equation}\label{barg-1}
\barg^{-1}(\vec{k};z,z')=\big [-\partial^{2}_{z} + {\vec{k}}^{2}+ M^2 
-\frac{3}{2} M^2 \sech^{2}\theta\big ]\delta(z-z')\; ,
\end{equation}
with the $\theta(z)$ previously defined. In terms of $R$ and $\rho$, we have
\begin{equation}\label{sigmasc}
\Sigma_{sc}(R,\rho)=\big [ -\frac{3}{2} M^2 \sech^{2}\theta(R+\frac{\rho}{2})\big ]\delta(\rho)\; .
\end{equation} 
Integrating over the $\rho$-coordinate, i.e., taking the Fourier transform at $k_{\rho}=0$, yields a $R$-dependent potential $V_{sc}(R)= -(3/2) M^2 \sech^{2}\theta(R)$, that vanishes at infinity. Therefore, the lowest order contribution to $\gatil^{(2)}(R;k_{\rho},\vec{k})$ at zero relative momenta is given by the mass squared plus a fluctuation potential that vanishes as $R\to\infty$. The first-order contribution can be split likewise: the constant term as $R\to\infty$ can be interpreted graphically as coming from the diagrams of Figure \ref{figautoenergia} computed at zero relative momenta with vacuum propagators, and with $\fith$ replaced by $\varphi_v$ in diagram $(b)$. Those are exactly the corrections to the mass (inverse bulk correlation length) squared. Our renormalization condition \eqref{gammaR(1)=0,gammaaR(2)=M2} guarantees that the constant term is just the renormalized mass squared; on the other hand, the contribution to the fluctuation potential, which vanishes as $R\to\infty$, is given by $\Delta \widetilde{\Sigma}(R;0,\vec{0})$. The fluctuation potential reflects the presence of the interface. 

\begin{figure}[h]
\vspace{0.5cm}
\centerline{\epsfig{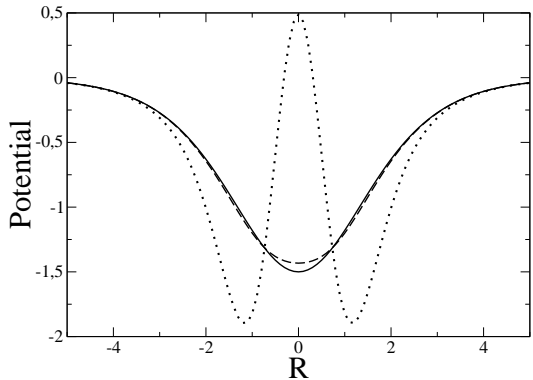}}
\caption{Potential as a function of $R$ for $\mu/M=0.1$ and $\lambda=0.1$. The solid line stands for $V_{sc}(R)$; the other lines correspond to the modified potential in $d=3$ (dashed) and $d=2$ (dotted).}\label{potentials}
\vspace{5mm}
\end{figure}

Using our propagator to compute the diagrams, we obtain for diagram $(a)$ of Figure \ref{figautoenergia}
\begin{subequations}
\begin{align}
\frac{\lambda}{24}\,\sech^2 \theta \bigg(\sqrt{3} \tanh^2 \theta +\frac{9}{4}\bigg(\frac{M}{\mu}-\frac{2}{\pi}\bigg)\,\sech^2 \theta \bigg), \;\;\; d=2\;\; ,\\
\frac{3\lambda M}{32\pi}\,\sech^2 \theta \bigg((\log 3)\, \tanh^2 \theta +\log \bigg(\frac{2M}{\mu}\bigg )\,\sech^2 \theta \bigg), \;\;\; d=3\;\; .
\end{align}
\end{subequations}
The calculations were done using Mathematica. The integrations for diagram $(b)$ of Figure \ref{figautoenergia} had to be performed numerically, so that we have computed it for several values of $R$ in order to draw the curve of the modified potential illustrated in Figure \ref{potentials}. As before, we have adopted the renormalization conditions in \eqref{gammaR(1)=0,gammaaR(2)=M2}. The modified potential is very sensitive to the value of the ratio $\mu/M$, especially in the case $d=2$. A comparison between Figures \ref{potentials} and \ref{profiled2} suggests that it is probably easier to measure the effects of fluctuations on the potential than on the interface profile.

Radiation scattered through the interface may be used to probe structure factors, which ultimately measure the two-point function. A numerical evaluation of $\gtil^{(2)}(R;k_{\rho},\vec{k})$ would then allow a direct comparison with data from binary liquid mixtures, or from more recent applications, especially those involving Bose-Einstein condensates. As we have said before, that calculation is feasible, but rather involved. Since $\gatil^{(2)}(R;k_{\rho},\vec{k})$ measures the change in free energy (which coincides with the effective action to first-order) as the profile changes, one might hope to have a direct test of our computation by measuring those changes for well-separated points at different values of $R$. Alternatively, Monte Carlo computations might be used as a test. 

\section{Conclusions}

We have cast all previous results for the correlation functions of interface fluctuations in the unified framework of a semiclassical expansion. Besides making contact with the existing literature, we have extended previous results to a lower dimension, making use of the closed analytic form for the semiclassical propagator. We have also computed susceptibilities and two-point correlations, which may eventually be checked experimentally.

It is important to note that our technique for resumming the spectral representation for the semiclassical propagator might be of use in other contexts, as long as we can reduce the differential equation for the propagator to an ordinary one. In the present case, an additional simplification came from the fact that the hypergeometric series involved terminated. In particular, spherically symmetric backgrounds would be natural candidates to be investigated.

We should emphasize that with our basic ingredient, the semiclassical propagator, we have reduced the calculation of physical quantities to computing Feynman diagrams, whose propagators and vertices carry information about the background solution. 

Many systems of interest may profit from the semiclassical treatment that we have presented. Thus, computing correlations in supersymmetric models is clearly a direction for future work. Likewise, correlations for mixtures of Bose-Einstein condensates separated by an interface are certainly worth pursuing. For this latter example, existing experimental techniques\cite{bnu} may open up a host of possibilities for experimental tests and checks. Also promising are the possibilities of comparison with experimental data in the more traditional binary liquid mixtures. We should emphasize that our treatment allows for a complete and separate treatment of capillary (those in the zero-mode subspace) and noncapillary waves. Reflectivities and form factors extracted from scattered radiation are the physical quantities to be measured for comparison. Monte Carlo simulations might also be used as a test.
 
Finally, we may hope to calculate other correlations of interest to experimentalists, as long as their defining Feynman diagrams lead to tractable integrals. As mentioned before, the two-loop calculation done previously for the surface tension may serve as a test of the simplification introduced by our expression for the propagator.

\acknowledgements

The authors acknowledge support from CAPES, CNPq, FAPERJ, and FUJB/UFRJ.


\appendix{}

\section{}
\label{The semiclassical propagator}

The homogeneous version of Equation \eqref{DiffG(k;z,z')=delta} in the text can be written in terms of dimensionless variables
\begin{equation}
\big [-\partial_{\theta}^{2} + b^2 -6\,\sech^2 \theta \big ]\,\phi \,=\,0\hspace{5mm},
\end{equation}
where $u \equiv (1-\tanh \theta)/2$, as before. Defining $\phi \equiv (\cosh \theta)^{-b}\,F(u)$, the function $F$ satisfies a hypergeometric equation
\begin{equation}
u(1-u)\,\frac{d^{2}F}{du^2} + \big [(b+1)-2(b+1)u \big ]\,\frac{dF}{du} + \big [6-b(b+1) \big ]\,=\,0\;\;\ ,
\end{equation}
whose general solution is
\begin{equation}\label{solucaoF=hypergeometric}
F(u) = c_{1}\,\big [\,{}_{2}F_{1} \big(b-2, b+3; 1+b;u \big)\big] +c_{2}\,\big [ u^{-b}\,{}_{2}F_{1} \big(3, -2; 1-b;u \big)\big ] \;\; ,
\end{equation}
where ${}_{2}F_{1}(A,B;C;u)$ is a hypergeometric function. The identity
\begin{equation}
{}_{2}F_{1}(A,B;C;u) = (1-u)^{C-A-B}\,{}_{2}F_{1}(C-A,C-B;C;u)\;\; ,
\end{equation}
yields
\begin{equation}\label{solucaoF=hypergeometric}
F(u) = c_{1}\,(1-u)^{-b}\,\big [\,{}_{2}F_{1} \big(3, -2; 1+b;u \big)\big ] +c_{2}\,u^{-b}\,\big [\,{}_{2}F_{1} \big(3, -2; 1-b;u \big)\big ] \;\; .
\end{equation}
Both series terminate. The two solutions in the linear combination of \eqref{solucaoF=hypergeometric} correspond to $\phi(b,u)$ and $\phi(-b,u)$ of equation \eqref{fi(b,u)=...} of the text.

{}From the solutions of the homogeneous equation, one constructs the semiclassical propagator by a standard procedure\cite{aragao=TQFDW}, which leads to \eqref{propsc(b,u)}. That expression can be rewritten as
\small
\begin{equation}
\barg = \frac{e^{-\sqrt{\vec{k}^2+M^2}\,(z-z')}}{2\sqrt{\vec{k}^2+M^2}}\,f(b,u)\,f(-b,u')\,\Theta(z-z')\;+\;
\frac{e^{-\sqrt{\vec{k}^2+M^2}\,(z'-z)}}{2\sqrt{\vec{k}^2+M^2}}\,f(-b,u)\,f(b,u')\,\Theta(z'-z)\;.
\end{equation}
\normalsize
The latter expression appears as one of the terms of the integral
\begin{equation}
{\cal I} \equiv \int_{-\infty}^{\infty}\frac{dq}{2\pi}\,\frac{\psi(q,z)\psi^{*}(q,z')}{q^2+\vec{k}^2+M^2}\hspace{5mm},
\end{equation}
where $\psi(q,z) \equiv e^{iqz}\,f(\frac{-2iq}{M},u)$. Indeed, its $\Theta(z-z')$ part has poles in the upper-half of the complex $q$-plane at $i\sqrt{\vec{k}^2+M^2}$, $-iM$ and $+iM$, with residues given by
\begin{subequations}
\begin{align}
&e^{-\sqrt{\vec{k}^2+M^2}(z-z')}\,f(b,u)f(-b,u')/4\pi i\sqrt{\vec{k}^2+M^2}\;,\\
&-12M\,u(1-u)u'(1-u')/4\pi i\vec{k}^2\; ,\\
&-6M\sqrt{u(1-u)}(1-2u) \,\sqrt{u'(1-u')}\,(1-2u')/4\pi i(\vec{k}^2+\frac{3M^2}{4})\,,
\end{align}
\end{subequations}
respectively. The $\Theta(z-z')$ part has similar contributions. Combining them, we derive
\begin{equation}
{\cal I} = \barg(\vec{k};z,z') -\bigg (\frac{M}{2}\bigg) \,\frac{\bar{\eta}_{0}(z)\bar{\eta}_{0}(z')}{\vec{k}^2}-
\bigg (\frac{M}{2}\bigg) \,\frac{\bar{\eta}_{1}(z)\bar{\eta}_{1}(z')}{\vec{k}^2+\frac{3M^2}{4}} \hspace{5mm},
\end{equation}
which leads to the spectral representation for $\barg$
\begin{equation}
\barg= \frac{M}{2}\bigg \{\frac{\bar{\eta}_{0}(z)\bar{\eta}_{0}(z')}{\vec{k}^2}+
\frac{\bar{\eta}_{1}(z)\bar{\eta}_{1}(z')}{\vec{k}^2+\frac{3M^2}{4}}\bigg \}+
\int_{-\infty}^{\infty}\frac{dq}{2\pi}\,\frac{\psi(q,z)\psi^{*}(q,z')}{q^2+\vec{k}^2+M^2}  \hspace{5mm},
\end{equation}
where one clearly identifies the eigenvalues and the eigenfunctions for the P\"oschl-Teller problem at hand. Using the variable $u$, we have
\begin{equation}
\begin{array}{ll}
\lambda_{0} = \vec{k}^2\;\; ;\;\; & \bar{\eta}_{0}(u)=2\sqrt{3}\,u(1-u)\;\;\; ,\\
\lambda_{1} = \vec{k}^2+\frac{3M^2}{4}\;\; ;\;\; & \bar{\eta}_{1}(u)=\sqrt{6u(1-u)}\,(1-2u)\;\;\; ,\\
\lambda_{q} = q^2+\vec{k}^2+M^2\;\; ;\;\; & \bar{\eta}_{q}(u)=e^{iqz}\,f\big(-2iq/M,u\big)\;\;\; .
\end{array}
\end{equation}

Having shown that our semiclassical propagator does correspond to the usual spectral representation, we now investigate its limit when $\kappa \rightarrow 0\,(b \rightarrow 2)$ in the subspace orthogonal to the zero eigenmode. That limit is required for the calculations of Section V. Unfortunately, the expression for the limit which appeared in reference \cite{aragao=TQFDW} is wrong (it is not orthogonal to the zero-mode subspace). We profit from this occasion to exhibit the correct expression
\begin{align}
\begin{split}
\barg(\vec{0};u,u') = \frac{2}{M}\,\bigg \{\frac{u(1-u')}{4u'(1-u)}\,{\cal G}(u, u') + \frac{3}{2}\,u(1-u)u'(1-u')\,\log\bigg [ \frac{u(1-u')}{u'(1-u)}\bigg ] - \\
-\frac{11}{2}\,u(1-u)u'(1-u') \bigg \}\, \Theta(u'-u) \;\;\; + \;\;\; (u\, \leftrightarrow u')\, ,
\end{split}
\end{align}
where
\begin{equation}
{\cal G}(u,u') = (1-u)^2+6u'(1-u)^2+6{u'}^{2}(1-u) + {u'}^2\;.
\end{equation}
The corrected expression for $\barg$ is indeed orthogonal to the zero mode subspace, as can be verified in a straightforward calculation.

\section{}
\label{The Renormalization Procedure}

In this Appendix, we shall outline the renormalization procedure adopted  in the text. We start from the effective action functional, up to one-loop order
\begin{equation}
{\cal A}[\phi(x)] \equiv \lambda \,\Gamma[\phi(x)] \,=\, S[\phi(x)] + \frac{\lambda}{2}\, \log \bigg (\frac{\Delta \big[\phi(x)\big ]}{\Delta_{v}} \bigg )\;\; ,
\end{equation}
written in terms of renormalized parameters, and add to it counterterms, in order to obtain a renormalized expression
\begin{align}
\begin{split}
{\cal A}_{R}\big [\phi(x) \big ] = {\cal A}\big [\phi(x) \big ] &- \frac{C_{1}}{2}\int d^d x \big (\phi^2 -\varphi_{v}^{2}\big ) - \frac{C_{2}}{4}\int d^d x \big (\phi^2 -\varphi_{v}^{2}\big )^2 \\
&-\frac{C_{3L}}{2}\int d^d x \big (\partial_{L}\phi \big )^2 -\frac{C_{3T}}{2}\int d^d x \big (\nabla_{T}\phi \big)^2\;\; .  
\end{split}\label{gammaR[fi(x)]}
\end{align}
In the formulae above, $\phi(x) \equiv \bra \varphi(x)\ket $ is the expectation value of the field. The renormalization constants $C_{1}, C_{2}, C_{3L}$ and $C_{3T}$ are associated to mass, coupling, and longitudinal and transverse wave function renormalization. They will be fixed by renormalization conditions in the vacuum sector. 

Functional derivatives of \eqref{gammaR[fi(x)]} with respect to $\phi(x)$ lead to the $n$-point vertex functions. Derivatives taken at $\phi(x)= \fith(x)$ yield vertices in the kink sector. As $\fith$ satisfies the equation of motion, for the renormalized one-point function in the {\em kink sector}, we obtain
\begin{align}\label{GammaR(1)[fichapeu]}
\Gamma_{R}^{(1)}[\fith] \,=\, \frac{\lambda}{2} \fith G - C_{1}\fith -C_{2}\fith \big (\fith^2-\varphi_{v}^2\big) + C_{3L} \big (\partial_{z}\fith\big)^2\;\; ,
\end{align}
where all quantities are taken at a given point $x$, so that $G = G(x,x)$ is the semiclassical propagator at coincident points. For the two-point vertex, we have
\begin{align}
\begin{split}\label{GammaR(2)[fichapeu;x1,x2]}
\Gamma_{R}^{(2)}[\fith; x_{1},x_{2}] \,=\, &G^{-1}(x_{1},x_{2}) + \frac{\lambda}{2} \delta^{(d)}(x_{1}-x_{2}) \, G - \frac{\lambda}{2} \fith(x_{1}) G(x_{1},x_{2}) \fith(x_{2}) G(x_{2},x_{1}) \\
&- \big [\,C_{1}+C_{2}\big (3\fith^2-\varphi_{v}^2\big) - C_{3L} \partial_{z}^2 - C_{3T}\nabla_{T}^{2}\, \big ] \, \delta^{(d)}(x_{1}-x_{2})\;\; ,
\end{split}
\end{align}
where again $\fith = \fith(x_{1})$, and $G=G(x_{1},x_{1})$.

Expressions for the {\em vacuum sector} are obtained from functional derivatives at $\varphi_{v}$. In the formulae above, that amounts to replacing $\fith$ and $G$ with $\varphi_{v}$ and $G_{v}$, respectively. One may derive expressions for $\Gamma_{R}^{(3)}$ and $\Gamma_{R}^{(4)}$, as well. The translational invariance in the vacuum sector makes it convenient to go to momentum space. Furthermore, we shall adopt zero-momentum renormalization conditions. If we define the Fourier transformed vertices as
\begin{equation}
\gatil^{(n)}(p_{1}, \ldots, p_{n}) \equiv (2\pi)^d \, \delta^{(d)}\bigg (\sum_{i}p_{i} \bigg )\, \barga^{(n)}(p_{1},\ldots,p_{n-1})\, ,
\end{equation}
we arrive at the following relations at zero-momenta
\begin{subequations}\label{renormalizedGammasAppB}
\begin{align}
&\barga_{R}^{(1)} = \frac{\lambda}{2} \varphi_{v} \int \gtil_{v}({k}) - C_{1}\varphi_{v} \;\; ,\\
&\barga_{R}^{(2)}({p}_{i}={0}) = M^2+ \frac{\lambda}{2} \int \gtil_{v}({k}) - \frac{\lambda \varphi_{v}^{2}}{2} \int \gtil_{v}({k}) - C_{1} - 2C_{2}\varphi_{v}^{2} \;\; ,\\
&\bigg (\frac{\partial \barga_{R}^{(2)}}{\partial [p_{z}^{2}]} \bigg ) = 1 - \frac{\lambda \varphi_{v}^{2}}{2}\frac{\partial }{\partial [p_{z}^{2}]} \bigg [\int \gtil_{v}({k})\gtil_{v}({k}+{p}) \bigg  ]_{{p}_{i}={0}} - C_{3L}\;\; ,\\
&\bigg (\frac{\partial \barga_{R}^{(2)}}{\partial [p_{T}^{2}]} \bigg ) = 1 - \frac{\lambda \varphi_{v}^{2}}{2}\frac{\partial }{\partial [p_{T}^{2}]} \bigg [\int \gtil_{v}({k})\gtil_{v}({k}+{p}) \bigg  ]_{{p}_{i}={0}} - C_{3T}\;\; ,\\
&\barga_{R}^{(4)}({p}_{i}={0}) = \lambda - \frac{3\lambda}{2} \int \gtil_{v}^{2}({k}) + 6\lambda \varphi_{v}^{2} \int \gtil_{v}^{3}({k}) - 3\lambda \varphi_{v}^{4} \int \gtil_{v}^{4}({k}) - 6C_{2}\;\;,
\end{align}
\end{subequations}
where all integrals are calculated with a cutoff in transverse momentum space. 
If we define
\begin{equation}
I_{n}(d,\Lambda) = \int_{\Lambda}\frac{d^{d-1} k_{T}}{(2\pi)^{d-1}}\int_{-\infty}^{\infty}\frac{dk_{L}}{2\pi} \, \gtil_{v}^{\,n}(k)\;\;
\end{equation}  
($k_{T}$ and $k_{L}$ amount to transverse and longitudinal momentum, respectively) relations \eqref{renormalizedGammasAppB} become
\begin{subequations}
\begin{align}
&\barga_{R}^{(1)} = \frac{\lambda}{2} \varphi_{v} I_{1}(d,\Lambda) - C_{1}\varphi_{v} \;\; ,\\
&\barga_{R}^{(2)}({p}_{i}={0}) = M^2+ \frac{\lambda}{2} I_{1}(d,\Lambda) - \frac{\lambda \varphi_{v}^{2}}{2} I_{2}(d,\Lambda) - C_{1} - 2C_{2}\varphi_{v}^{2} \;\; ,\\
&\bigg (\frac{\partial \barga_{R}^{(2)}}{\partial [p_{z}^{2}]} \bigg ) = 1 - \frac{\lambda \varphi_{v}^{2}}{2} \bigg [3 I_{3}(d,\Lambda) -4M^2 I_{4}(d,\Lambda) - \frac{16 \pi^2 \Omega_{d}}{\Omega_{d+2}} I_{4}(d+2,\Lambda)\bigg ]- C_{3L}\;\; ,\\
&\bigg (\frac{\partial \barga_{R}^{(2)}}{\partial [p_{T}^{2}]} \bigg ) = 1 - \frac{\lambda \varphi_{v}^{2}}{2} \bigg [I_{3}(d,\Lambda) - \frac{16 \pi^2 \Omega_{d}}{d \Omega_{d+2}} I_{4}(d+2,\Lambda)\bigg ] - C_{3T}\;\; ,\\
&\barga_{R}^{(4)}({p}_{i}={0}) = \lambda - \frac{3\lambda}{2} I_{2}(d,\Lambda) + 6\lambda \varphi_{v}^{2} I_{3}(d,\Lambda) - 3\lambda \varphi_{v}^{4} I_{4}(d,\Lambda) - 6C_{2}\;\; ,
\end{align}
\end{subequations}
where $\Omega_{d}$ is the usual $d$-dimensional solid angle. 

The renormalization conditions \eqref{gammaR(1)=0,gammaaR(2)=M2} for the vacuum sector which were adopted in the text lead to the determination of the constants: $C_{1}= \lambda I_{1}(d,\Lambda)/2$, $C_{2}= -\lambda I_{2}(d,\Lambda)/4$, $C_{3L}=C_{3T}=0$. Using such values in the $\Lambda$-regulated Fourier transformed expressions \eqref{GammaR(1)[fichapeu]} and \eqref{GammaR(2)[fichapeu;x1,x2]} at zero momenta cancels the ultraviolet divergences as $\Lambda \rightarrow \infty$.

As a consistency check on our procedure, we have used the renormalization conditions
\begin{equation}
\barga_{R}^{(1)}=0\, ; \;\;\;\;\barga_{R}^{(2)}=M^2\, ;\;\;\;\; \bigg (\frac{\partial \barga_{R}^{(2)}}{\partial [p_{L}^{2}] } \bigg )_{{0}} = 1\;\; ,
\end{equation} 
that were adopted in reference \cite{peter}, in order to calculate corrections to the kink mass. The results we have found coincide with those of \cite{peter}, illustrating that the calculation of the determinant via the semiclassical propagator is not afflicted with mode-counting ambiguities. That allows us to use a simple momentum cutoff regularization, which is more directly related to condensed matter phenomenology.


\end{document}